\documentclass[]{spie}  %>>> use for US letter paper
\usepackage{amsmath,amsfonts,amssymb}
\usepackage{graphicx}
\usepackage[colorlinks=true, allcolors=blue]{hyperref}

\title{CONCERTO: a breakthrough in wide field-of-view spectroscopy at millimeter wavelengths}

\author[a,b]{Alessandro~Fasano}
\author[a]{Alexandre~Beelen}
\author[c,b]{Alain~Beno\^it}
\author[d]{Andreas~Lundgren}
\author[e]{Peter~Ade}
\author[g]{Manuel~Aravena}
\author[c,b]{Emilio~Barria}
\author[a]{Matthieu~B\'ethermin}
\author[g,b]{Julien~Bounmy}
\author[g,b]{Olivier~Bourrion}
\author[c,b]{Guillaume~Bres}
\author[c,b]{Martino~Calvo}
\author[c,b]{Andrea~Catalano}
\author[h,b]{François-Xavier~D\'esert}
\author[d]{Carlos~De Breuck}
\author[i]{Carlos~Dur\'an}
\author[a]{Thomas~Fenouillet}
\author[a]{Jose~Garcia}
\author[c,b]{Gregory~Garde}
\author[c,b]{Johannes~Goupy}
\author[j]{Christopher~Groppi}
\author[g,b]{Christophe~Hoarau}
\author[a]{Wenkai~Hu}
\author[a]{Guilaine~Lagache}
\author[a]{Jean-Charles~Lambert}
\author[c,b]{Jean-Paul~Leggeri}
\author[c,b]{Florence~Levy-Bertrand}
\author[g,b]{Juan-Francisco~Mac\'ias-P\'erez}
\author[j]{Hamdi~Mani}
\author[g,b]{Julien~Marpaud}
\author[j]{Philip~Mauskopf}
\author[c,b]{Alessandro~Monfardini}
\author[c]{Giampaolo~Pisano}
\author[h,b]{Nicolas~Ponthieu}
\author[a]{Leo~Prieur}
\author[g,b]{Samuel~Roni}
\author[g,b]{Sebastien~Roudier}
\author[g,b]{Damien~Tourres}
\author[e]{Carol~Tucker}
\author[a]{Mathilde~Van~Cuyck}

\affil[a]{Aix Marseille Univ., CNRS, CNES, LAM, Marseille, France}
\affil[b]{Groupement d'Interet Scientifique KID, 38000 Grenoble and 38400 Saint Martin d'H\'eres, France}
\affil[c]{Univ. Grenoble Alpes, CNRS, Grenoble INP, Institut N\'eel, 38000 Grenoble, France}
\affil[d]{European Southern Observatory, Karl Schwarzschild Straße 2, 85748 Garching, Germany}
\affil[e]{Astronomy Instrumentation Group, University of Cardiff, The Parade, CF24 3AA, United Kindgom}
\affil[f]{N\'ucleo de Astronom\'ia, Facultad de Ingenier\'ia y Ciencias, Universidad Diego Portales, Av.  Ej\'ercito 441, Santiago, Chile}
\affil[g]{Univ. Grenoble Alpes, CNRS, LPSC/IN2P3, 38000 Grenoble, France}
\affil[h]{Univ. Grenoble Alpes, CNRS, IPAG, 38400 Saint Martin d'H\'eres, France}
\affil[i]{European Southern Observatory, Alonso de Cordova 3107, Vitacura, Santiago, Chile}
\affil[j]{School of Earth and Space Exploration and Department of Physics, Arizona State University, Tempe, AZ 85287, USA}

\authorinfo{Further author information: (Send correspondence to Alessandro Fasano)\\Alessandro Fasano: E-mail: alessandro.fasano@lam.fr}

\begin{document} 
\maketitle

\begin{abstract}
CarbON CII line in post-rEionization and ReionizaTiOn (CONCERTO) is a low-resolution spectrometer with an instantaneous field-of-view of 18.6\,arcmin, operating in the 130--310\,GHz transparent atmospheric window. It is installed on the 12-meter Atacama Pathfinder Experiment (APEX) telescope at 5\,100\,m above sea level. The Fourier transform spectrometer (FTS) contains two focal planes hosting a total of 4\,304 kinetic inductance detectors. The FTS interferometric pattern is recorded on the fly while continuously scanning the sky. One of the goals of CONCERTO is to characterize the large-scale structure of the Universe by observing the integrated emission from unresolved galaxies. This methodology is an innovative technique and is called line intensity mapping. In this paper, we describe the CONCERTO instrument, the effect of the  vibration of the FTS beamsplitter, and the status of the CONCERTO main survey.
\end{abstract}


\section{INTRODUCTION}
\label{sec:intro}  % \label{} allows reference to this section
The CarbON CII line in post-rEionization and ReionizaTiOn (CONCERTO) instrument has been designed to provide a breakthrough in wide field-of-view (FoV) spectroscopy at millimeter (mm) wavelengths with its large mapping speed and multi-frequency capabilities. The primary scientific target is the observation of [CII]-emission line at redshift $z\ge$5.2 using line intensity mapping (LIM). This is also one of the main goals of CCAT-prime\cite{choi2020}
and TIME \cite{crites2014}.
Line intensity mapping consists of retrieving the 3-D information of the line intensity fluctuations, by measuring angular fluctuations in the brightness of the sky, rather than seeing individual point sources. The time (third) dimension, beyond the two sky coordinates, is obtained by observing at several electromagnetic frequencies where the line is redshifted in correspondence with different epochs.
The [CII] line is among the brightest lines originating from star-forming  galaxies and it is a reliable  tracer  of  star  formation on global scale. The ultimate goal is to answer the questions of whether dusty star-formation contributes to early galaxy evolution, and whether dusty galaxies play an important role in reionization.
The Cosmic Evolution Survey (COSMOS) field is the selected region for the CONCERTO LIM survey \cite{2020A&A...642A..60C}. It has already been extensively observed both on visible-IR photometry and spectroscopy which will be highly beneficial for the foregrounds subtraction in the LIM \cite{yue2015}.

CONCERTO is based on two arrays for a total of 4\,304 kinetic inductance detectors (KIDs) coupled to a Fourier transform spectrometer (FTS) in the ``fast scanning'' technique\cite{2013MNRAS.429..849D} for interferometric pattern sampling: we record the spectrum on the fly (otf) while continuously mapping on-sky.

In this paper, we introduce the CONCERTO technical main characteristics, the effect of the FTS beamsplitter vibration, and the status of the [CII] survey.
In Sect.~\ref{sec:instrument}, we describe the CONCERTO instrument.
Section~\ref{sec:beamsplitter} presents a peculiar challenge of on-ground FTSs, the vibrations in the instrument have to be controlled and minimized.
Finally, Sect.~\ref{sec:results} presents the instrument mapping strategy and the status of the observations of the COSMOS field.

\section{CONCERTO INSTRUMENT}
\label{sec:instrument}

CONCERTO has been designed with the following constraints in mind\cite{2018IAUS..333..228L}: 1) the maximization of the sensitivity and the minimization of the detectors' time response, 2) the large coverage both on sky-area and electromagnetic band (130--310\,GHz), 3) the absolute spectral resolution down to 1.5\,GHz and the moderate angular resolution (central beam width of $\sim30$\,arcsec in the 195--310\,GHz band).
For such requirements, CONCERTO consists of a fast FTS and registers a full interferometric pattern of 70\,mm in length (forward and backward) at $\sim$2.48\,Hz ($\sim$0.40\,s) by exploiting the fast response of two arrays of lumped element KIDs \cite{2010SPIE.7741E..0MD}, granting a single point sampling at 3\,813\,Hz (0.26\,ms). Fourier transform spectroscopy consists of measuring the output intensity of an interferometer for a varying optical path difference (OPD) between two arms of the interferometer itself, which is typically achieved with a moving roof mirror\cite{fasano-ltd}. The interferogram is then analyzed by Fourier transform and the spectrum is obtained. In ground-based experiments, the fast sampling of the interferograms is crucial to deal with the atmospheric fluctuations and properly calibrate the signal\cite{fasano_aa}. 
In addition, FTS allows us to exploit a larger instantaneous FoV with respect to other spectrometric solutions (e.g., grating or Fabry-Perot) or dispersive elements (prism or dichroic filters).
In the FTS, the single pixel detects the whole electromagnetic signal and records a full interferogram. Rather than the single-pixel efficiency, the gain is obtained by adopting large format arrays, increasing the mapping speed. In other words, the FTS permits the exploitation of a single large array by multiplexing the electromagnetic frequencies.

In CONCERTO, a linear motor controls the moving mirror motion responsible for the OPD. The motor size exceeds 0.5\,m in length and 3\,kg in mass. A second, twin, motor counterbalances the linear momentum associated with the moving mass by oscillating with an opposite instantaneous velocity \cite{2020A&A...642A..60C}.

The CONCERTO spectrometer exhibits two orthogonal focal planes which host the frequency-overlapping high-frequency (HF, 195--310\,GHz) and the low-frequency (LF, 130--270\,GHz) arrays, respectively.
Among the possible FTS implementations, we adopted the Martin-Puplett interferometer (MPI) configuration\cite{mpi}.
The projection of the last MPI polarizer selects the two focal planes, which are in reflection (HF) and transmission (LF) with respect to the polarizer placed at 45\,deg\footnote{The polarizer is mechanically installed at 48\,deg with respect to the perpendicular axis of the instrument to take into account the wire grid projection.}.
Each of the twelve readout lines (six per array) is connected to a custom readout board, with a maximum multiplexing factor of 400 over a bandwidth of 1\,GHz\cite{2022JLTP..tmp...51M}.
The focal plane is cooled-down to 60\,mK (with stabilization of 0.1\,mK) by a custom $^3$He/$^4$He dilution refrigerator coupled to a Pulse-Tube for the higher-temperature stages. The CONCERTO dilution refrigerator is specially conceived to preserve its temperature stability at inclinations up to 75\,deg for the mixing chamber, allowing us to operate at telescope elevation comprised in the range of 15--90\,deg.

The instrumental characteristics of CONCERTO measured during the 2021-commissioning campaign on the sky are summarized in Table\,\ref{tab:specs}.

\begin{table}[ht]
\caption{Main characteristics of CONCERTO instrument coupled to the APEX telescope. \cite{2022EPJWC.25700010C} }
\label{tab:specs}
\begin{center}       
\begin{tabular}{|l|l|} 
\hline
\rule[-1ex]{0pt}{3.5ex}  Illuminated telescope primary mirror diameter & 11\,m  \\
\hline
\rule[-1ex]{0pt}{3.5ex}  Field-of-view diameter & 18.6\,arcmin   \\
\hline
\rule[-1ex]{0pt}{3.5ex}  Beam widths HF $\mid$ LF & $\sim$30 $\mid$ $\sim$35\,arcsec \\
\hline
\rule[-1ex]{0pt}{3.5ex}  Absolute spectral resolution & settable and down to $\sim$1.5\,GHz \\
\hline
\rule[-1ex]{0pt}{3.5ex}  Frequency range HF $\mid$ LF & 195--310 $\mid$ 130--270\,GHz  \\
\hline
\rule[-1ex]{0pt}{3.5ex}  Pixels on sky HF $\mid$ LF & 2\,152 $\mid$ 2\,152  \\
\hline
\rule[-1ex]{0pt}{3.5ex}  Instrument geometrical throughput & 2.5$\times$10$^{-3}$\,sr$\,$m$^2$  \\
\hline
\rule[-1ex]{0pt}{3.5ex} Single pixel geometrical throughput & 1.16$\times$10$^{-6}$\,sr$\,$m$^2$   \\
\hline
\rule[-1ex]{0pt}{3.5ex}  Data rate & 128\,MiBytes/sec  \\
\hline 
\end{tabular}
\end{center}
\end{table}

\section{MIRROR MOTION AND VIBRATIONS ON THE BEAMSPLITTER}
\label{sec:beamsplitter}

The precise sampling of the interferograms and the systematics control are two fundamental aspects that have to be taken into account in an FTS.
If on the one side each interferogram needs to be recorded rapidly (typically $<$1\,s\cite{2017A&A...599A..34R}), on the other side the demanding technological requirements are tricky to handle.
Concerning sampling speed, the acquisition rate is set by the need of properly sample the interferogram.
Concerning vibrations, both technological counter measurements and operational compromises have to be adopted: spiral springs, magnetic breaks, counter motors, and acceleration optimization on the motor\cite{fasano-ltd,2020A&A...642A..60C}.

\begin{figure} [ht]
   \begin{center}
   \begin{tabular}{c} 
   \includegraphics[height=6.5cm]{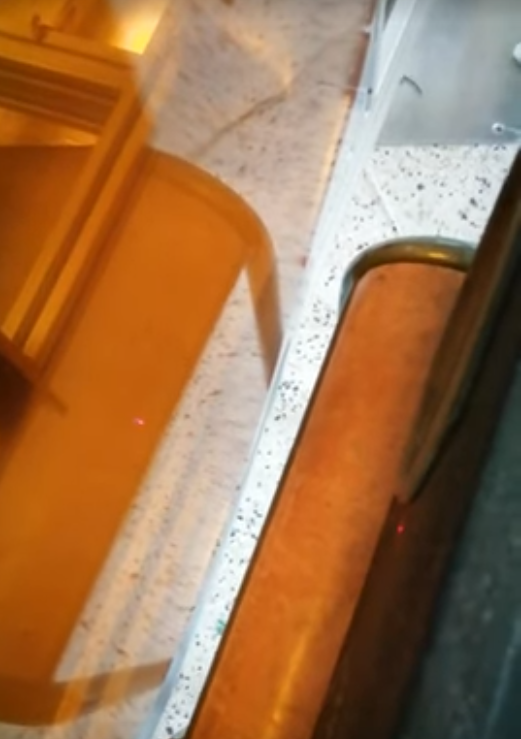}
	\end{tabular}
	\end{center}
   \caption{Laboratory characterization of the beamsplitter vibrations by creating a sound wave. The red laser pointing is visible in the center of the membrane. \url{https://www.youtube.com/shorts/s95272u9b1Y}}
   \label{fig:video} 
   \end{figure}

In CONCERTO, a peculiar effect interests the instrument: the roof mirror motion creates a wind flow that induces oscillations in the FTS beamsplitter with a resonance at $\sim$47.25\,Hz\footnote{This is the value of the resonance frequency for the specific polarizer used in CONCERTO.} causing a complex perturbation on the OPD. This is the consequence of the large beamsplitter area, a large (480$\times$800\,mm$^2$) elliptic 50\,$\mu$m-thick membrane of polyimide on which lie 50\,$\mu$m of copper wires with a pitch of 100\,$\mu$m. The video linked in Fig.~\ref{fig:video} shows this effect on a twin CONCERTO beamsplitter in the laboratory. This setup was conceived to reproduce the wind flow with the sound wave created by a speaker that makes vibrating the membrane.

Several solutions can be envisaged to mitigate and control this effect: 1) the use of a metallic wire-grid polarizer to replace the membrane-based beamsplitter, 2) the installation of a cross rigid support to attenuate and shift the oscillation at higher frequencies, 3) the real-time measurement of the OPD distortions, by directly measuring the vibrations on the beamsplitter, and 4) the development of an active feedback system capable to counter the oscillations with proper sound waves.
The first two solutions could not be adopted because, respectively: 1) the wire-grid polarizers are not produced with the CONCERTO size, and 2) mechanical support would reduce the optical efficiency.

\begin{figure} [ht]
   \begin{center}
   \begin{tabular}{cc} %% tabular useful for creating an array of images
   \includegraphics[height=6.5cm]{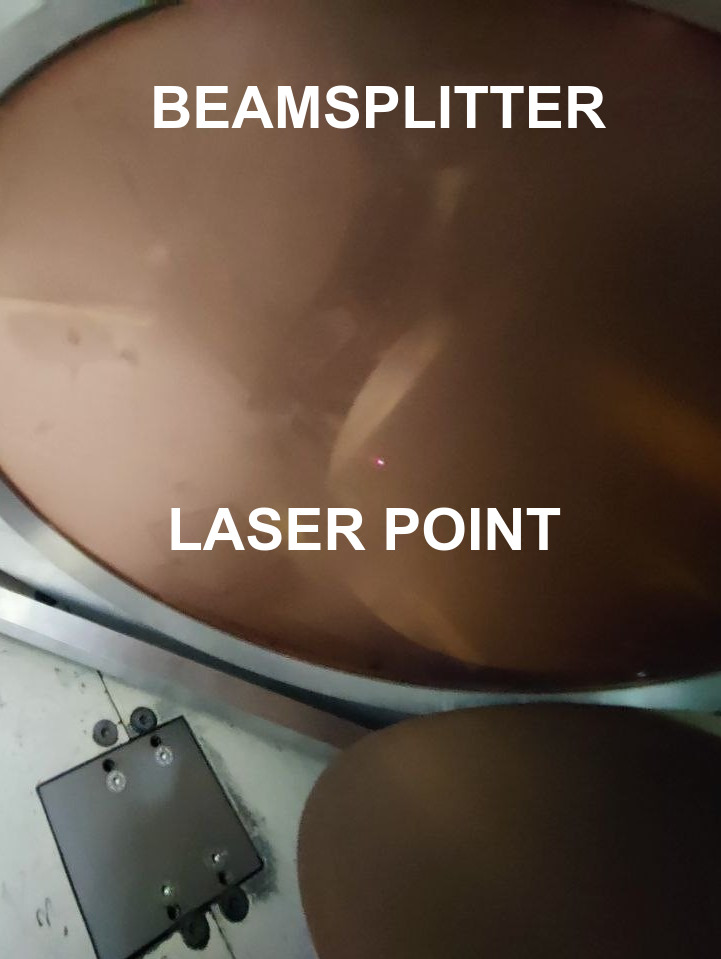}
   \includegraphics[height=6.5cm]{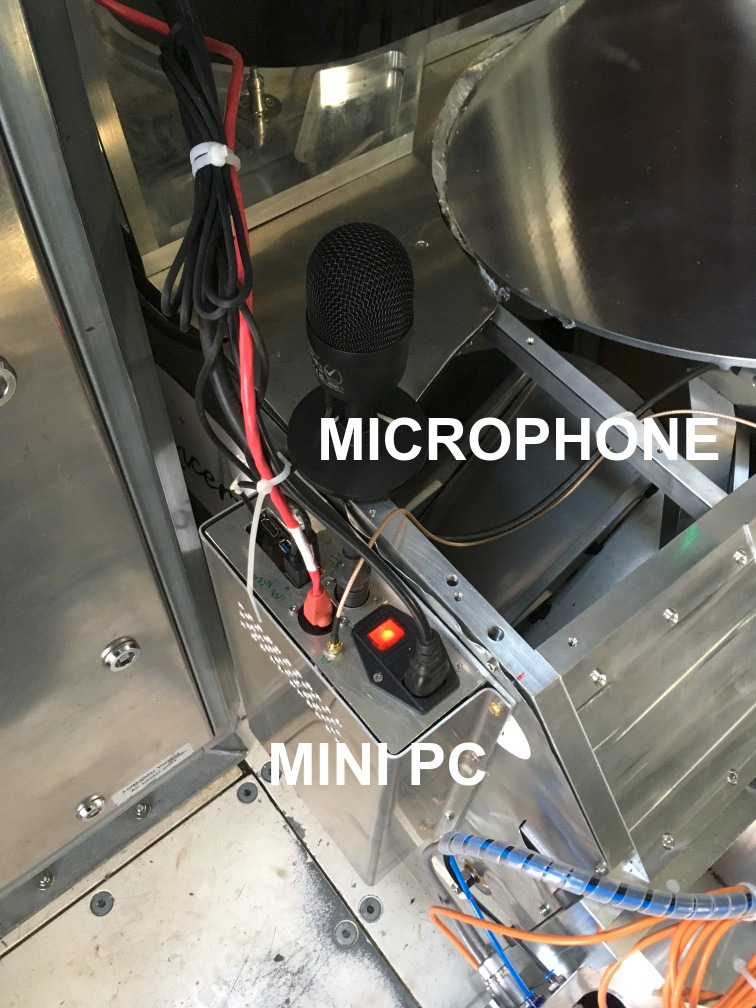}
   \includegraphics[height=6.5cm]{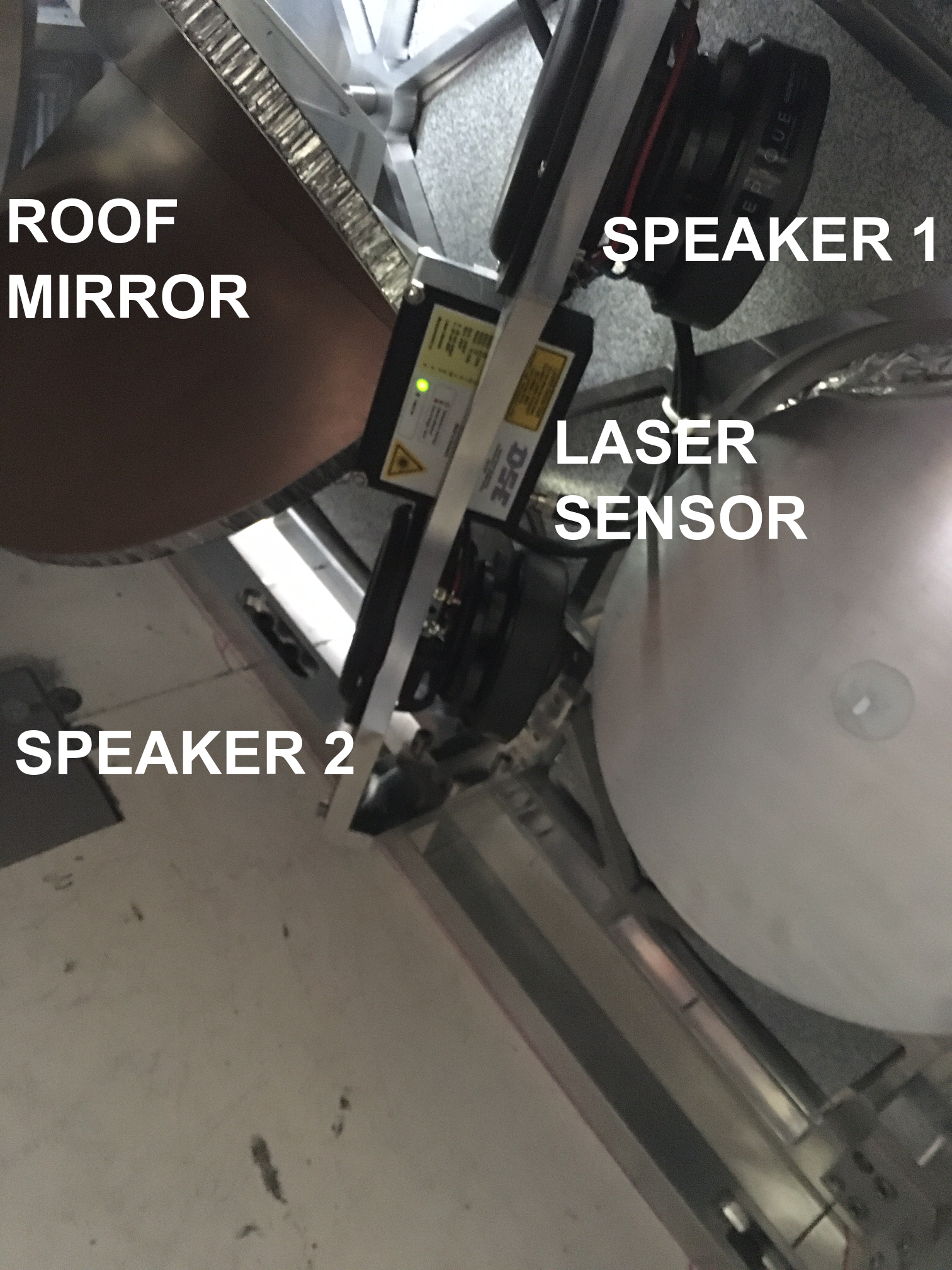}
   \end{tabular}
   \end{center}
   \caption{Counter feedback system hardware installed in the APEX Cassegrain cabin (C-cabin). Left: beamsplitter and red laser point in the center. Center: mini-pc and additional microphone to check the ambient noise in the C-cabin. Right: laser sensor and speakers.}
   \label{fig:counter}
\end{figure}

We implemented the last two solutions by installing a laser sensor pointing to the center of the beamsplitter in December 2021. This laser both measures the vibrations (around the nominal position) of the beamsplitter and allows us to calculate the related acoustic counter-wave to be applied for minimizing the oscillations.
The counter feedback system consists of a laser sensor, a dedicated mini-PC that calculates the counter-wave function, and two sound speakers as shown in Fig.~\ref{fig:counter}.

\begin{figure} [ht]
   \begin{center}
   \begin{tabular}{c} %% tabular useful for creating an array of images 
   \includegraphics[height=7cm]{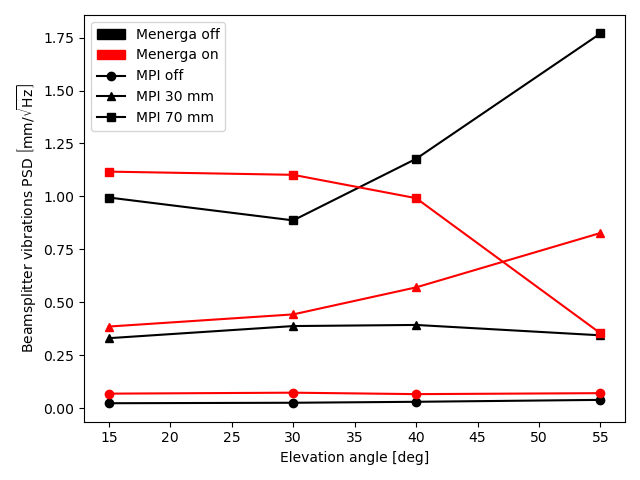}
   \end{tabular}
   \end{center}
   \caption{Power spectral density of the beamsplitter oscillation at 47.20\,Hz. Study at different MPI, elevation and Menerga cooling-system configurations. In black, the Menerga system is shut down; in red, the Menerga system is active. In dots, the MPI mirror is not moving; in triangles, the MPI spans 30\,mm; in squares, the MPI spans 70\,mm.}
   \label{fig:menerga}
\end{figure}

We performed several experiments to determine the cause of the beamsplitter vibrations and to characterize it in amplitude and frequency.
Firstly, we investigated the origin of the oscillation by studying its amplitude in different setups, as shown in Fig.~\ref{fig:menerga}. 
In this study, we adopted three different MPI configurations with respect to the roof mirror: completely stopped (off), spanning 30\,mm and 70\,mm at $\sim$2.48\,Hz ($\sim$0.40\,s). In the same experiment, we had explored an external cause for the beamsplitter oscillation, the Cassegrain cabin (C-cabin) wind flow produced by the fans of the air cooling system (so-called Menerga).
In addition, we pointed the telescope at different elevation angles to understand if there is a correlation with the oscillation amplitude. In principle, the beamsplitter is aligned with the rotation axis of the elevation, notwithstanding, a fraction of a degree in the misalignment in the angle might be tolerated.
The study suggests that the larger contribution of the beamsplitter vibrations comes from the MPI mirror, wherein in the 70\,mm case we reach a top-speed of $\sim$175\,mm/s and in the 30\,mm case $\sim$75\,mm/s. The contribution of the Menerga cooling system is minor compared to that visible from the MPI off case, where the power spectral density (PSD) slightly increases when the cooling system is activated (Menerga on).
On the other hand, no clear relation with the elevation angle is found as shown in Fig.~\ref{fig:menerga}.

\begin{figure} [ht]
   \begin{center}
   \begin{tabular}{cc} %% tabular useful for creating an array of images 
   \includegraphics[height=4.7cm]{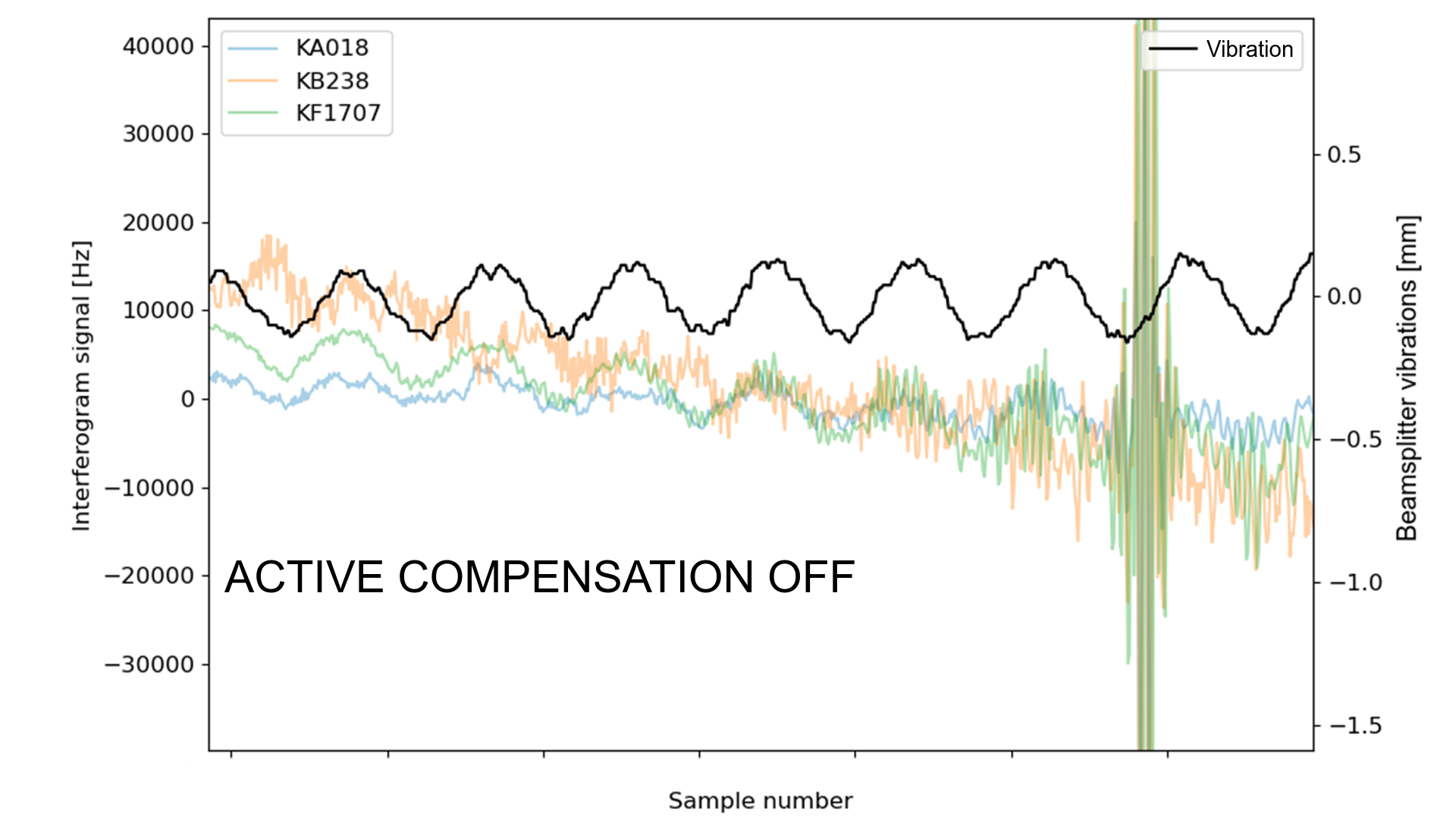}
   \includegraphics[height=4.7cm]{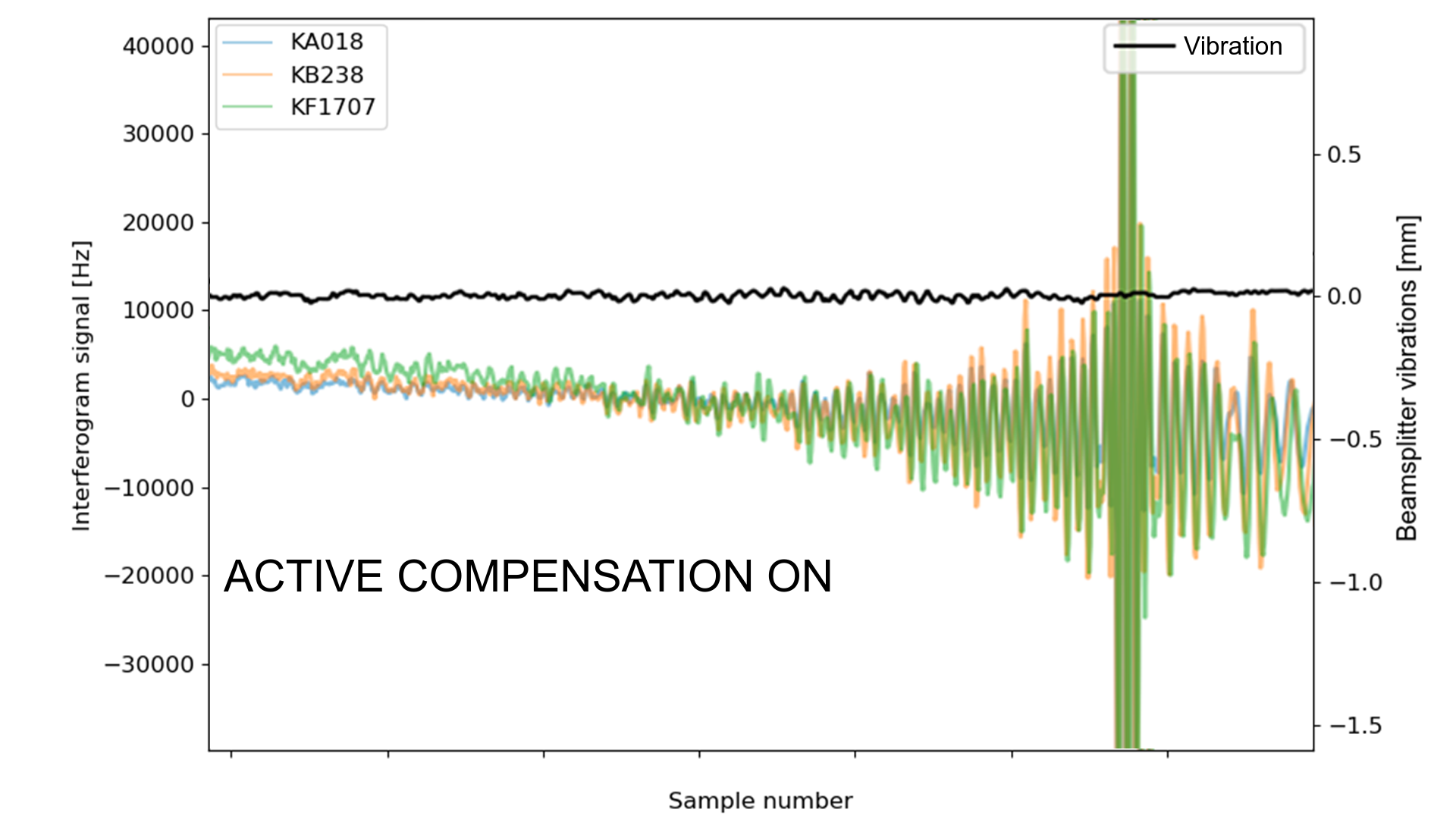}
   \end{tabular}
   \end{center}
   \caption{Time-ordered measurement of the beamsplitter oscillation (black) and three examples of interferogram signal (expressed in KID resonance-frequency shift [Hz]\cite{fasano_aa}). It is visible the presence of the interferogram patterns on the right side of both the figures. Each sample is recorded at 3\,813\,Hz (0.26\,ms). Left: the setup with counter feedback system off. Right: the evolution of the current instrument by adopting a slower MPI mirror speed ($\sim$0.54\,s spanning 70\,mm) and the counter feedback system on. The oscillation, as well as the related signal, is strongly reduced when the active compensation is on.}
   \label{fig:clean}
\end{figure}

Secondly, we studied the correlation between the oscillation and the time-ordered signal on the KIDs. As shown, in Fig.~\ref{fig:clean} on the left, the correlation is visible with the counter feedback system off.
When the counter feedback system is on, as shown in Fig.~\ref{fig:clean} on the right, it reduces the beamsplitter vibrations by a factor $>$7 (peak-peak).

\begin{figure} [ht]
   \begin{center}
   \begin{tabular}{c} %% tabular useful for creating an array of images 
   \includegraphics[height=7cm]{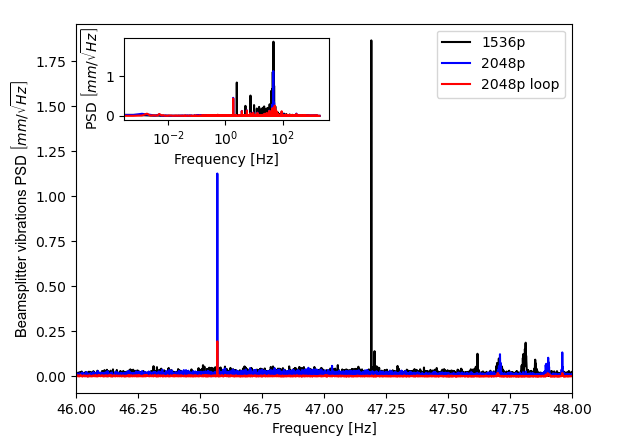}
   \end{tabular}
   \end{center}
   \caption 
   {Power spectral density (PSD) of the beamsplitter oscillation vs frequency. In black, the 1536-point sampling; in blue, the 2048-point one; in red, the 2048-point configuration with the counter feedback system active. The main figure displays the zoom on the frequency range of interest (46--48\,Hz), while the subfigure shows the whole frequency-domain range (up to 1906.5\,Hz for the Nyquist frequency).}
   \label{fig:laser3} 
\end{figure}

Finally, we characterized the oscillation in the frequency domain by studying its recorded signal by Fourier transform in different mirror-motion configurations, as shown in Fig.~\ref{fig:laser3}. 
While the spanning length of the roof mirror is fixed by the scientific requirements (70\,mm for the desired spectral resolution\cite{2020A&A...642A..60C}), the motor speed has some margins. Its motion curve is designed to quickly stabilize the mirror speed, by strongly accelerating it (up to 8\,m/s$^2$), and obtaining a homogeneous sampling of the interferogram.
In Figure~\ref{fig:laser3}, the number of sampling points corresponds to the number of points acquired during a whole forward/backward mirror motion. We show the fastest configuration with 1\,536 points ($\sim$0.40\,s) of sampling and the alternative one at 2\,048 points ($\sim$0.54\,s) with and without the counter feedback system. We adopted the lower speed to mitigate the wind flow for the new observations. At the same time, we better reconstruct the interferograms. As an additional effect, by adopting the 2048-point configuration and activating the counter feedback system, the $\sim$47.20\,Hz oscillation is shifted at a lower frequency ($46.50$\,Hz), as shown in Fig.~\ref{fig:laser3}.

\section{OBSERVATIONS STATUS}
\label{sec:results}

\begin{figure} [ht]
   \begin{center}
   \begin{tabular}{cc} %% tabular useful for creating an array of images 
   \includegraphics[height=6.25cm]{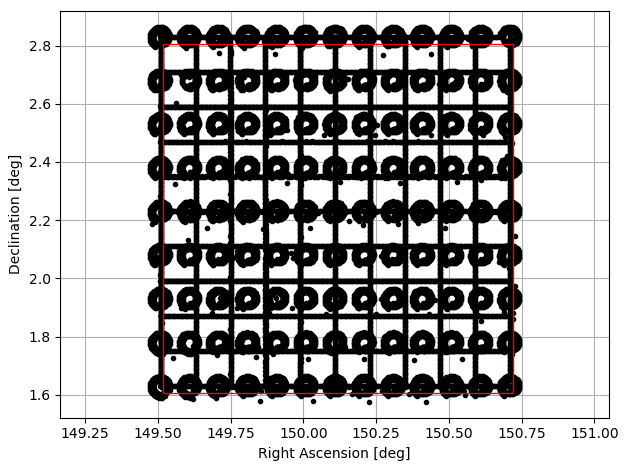}
   \includegraphics[height=6.25cm]{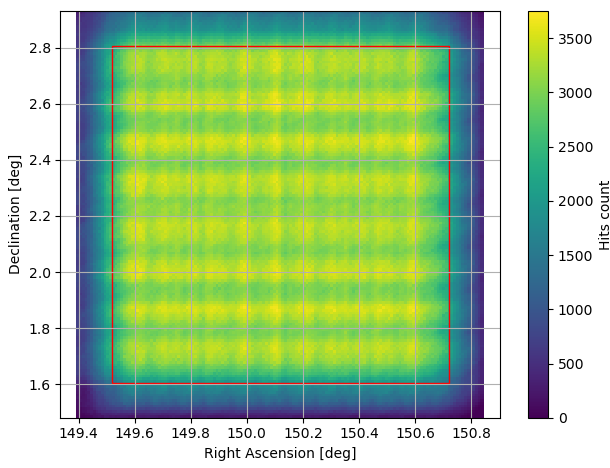}\\
   \includegraphics[height=6.25cm]{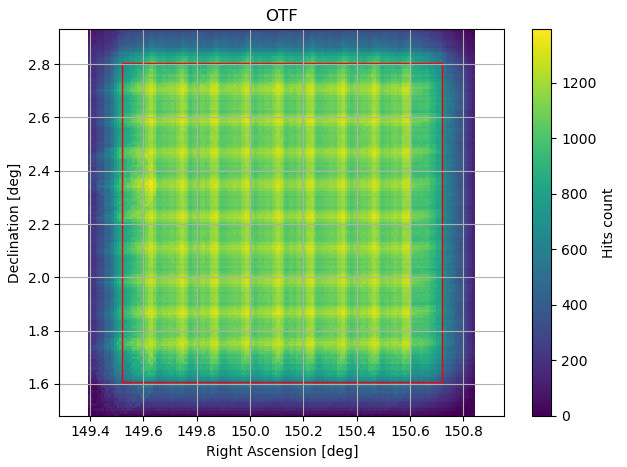}
   \includegraphics[height=6.25cm]{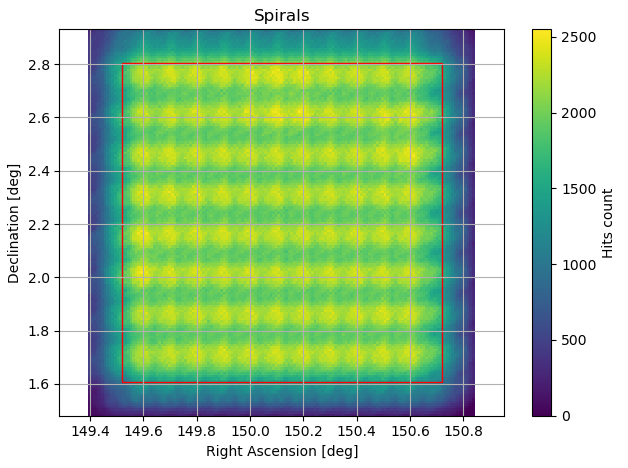}\\
   \end{tabular}
   \end{center}
   \caption{Sky coverage study, scans' number 22570, 22571 for the otf, and 22574--22577 for the spiral strokes. Scanning strategy employed from July 2021 to May 2022. 85\% of the identified KIDs (over a total of 3718) are present with eccentricity smaller than 0.8. In red is identified the area we are covering in the COSMOS field. The 2-d histograms are sampled with a squared bin of 35\,arcsec of edge (LF beam width). Up left: the projection of the telescope pointing in a whole COSMOS loop observations. Up right: the hit count of the sky-patch by considering a single array projected on the sky sampling at 0.53\,s. Bottom left: the separated hits count for the otfs. Bottom right: the separated hits count for the spiral strokes. }
   \label{fig:otf}
\end{figure}

The main scientific driver of the CONCERTO project is the observation of the [CII] line in the COSMOS\cite{2022ApJS..258...11W} field. This sky-patch of COSMOS covers a square of 72$\times72$\,arcmin$^2$ centered on 150.12\,deg right ascension and 2.21\,deg in declination.
The designed observing strategy takes into account the MPI necessity to sample a full interferogram to retrieve a whole electromagnetic spectrum before moving to the next spatial pixel. This sets a limit on the telescope scanning speed. Each point in the sky (defined by a beam width, i.e., 30\,arcsec for HF and 35\,arcsec for LF) has to be measured with, at least, a whole interferogram.
We sample a full double-interferogram (forward/backward) in a fraction of a second (0.40\,s or 0.54\,s, respectively for the 1\,536 and the 2\,048 points cases) and we fix the telescope maximum scanning speed at 40\,arcsec/s.
We have, thus, at least 2.13 full forward/backward mirror motions per beam width (a total of 4.16 interferograms)\footnote{This result is obtained by taking the worst scenario, the number of full mirror motions per beam width, N, being calculated following $ \rm{N} = 40\,\text{arcsec s$^{-1}$} / 30\,\rm{arcsec}/0.54\,\rm{s} \approx 2.13$ }.
In addition, the adopted mapping strategy has been conceived to reach the highest level of homogeneity of sky coverage.
Figure~\ref{fig:otf} shows the simulated and the observed complete COSMOS loop scanning, composed of two otf scans in raster scanning pattern and four raster scanning (step and integrate) with spiral strokes feature.
On the up-left figure, in vertical and horizontal black lines, we recognize the vertical and horizontal otf raster, respectively. Every shape similar to a thick circle (a spiral spanning a 4$\pi$-angle at low radial velocity) is a stroke of the point-and-integrate raster. The two different techniques are adopted to ensure a mix in the observations' angles and mitigate risks related to systematics induced by the scanning strategy. On-the-fly raster ensures a constant speed on the telescope scanning, while spiral strokes scanning involves high telescope speed when it reaches the end of the stroke.

A whole single loop requires $\sim$2.35\,hours and translates into 6 files for a total data amount of $\sim$1\,TBytes.

We observed a total of 314\,hours on the COSMOS project (including overheads) from July 2021 to May 2022, from the European Southern Observatory (ESO)\footnote{\url{https://www.eso.org}} and Onsala Space Observatory (OSO)\footnote{\url{https://www.chalmers.se/en/researchinfrastructure/oso/Pages/default.aspx}} allocated time and a further 55\,hours on Chilean allocated time. By stacking the maps for the 637 central KIDs in the HF array we obtain a measurement of the root mean squared (RMS) noise of the maps. Without relevant systemic errors, we expect that the RMS noise decreases as a function of the integration time (t), following a simple relation:

\begin{equation}
    \rm{RMS} (t) \propto \frac{1}{ t^\beta } \rm{,}
    \label{eq:rms}
\end{equation}

\noindent where $\beta$ is expected to be $\sim$0.5.

\begin{figure} [ht]
   \begin{center}
   \begin{tabular}{cc} %% tabular useful for creating an array of images 
   \includegraphics[height=6.cm]{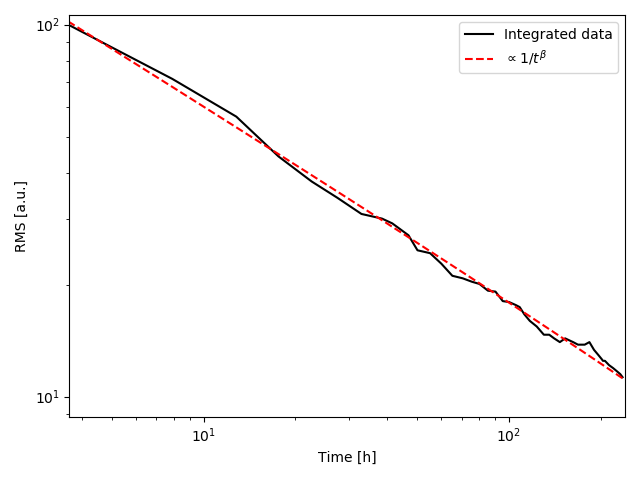}
   \includegraphics[height=6.cm]{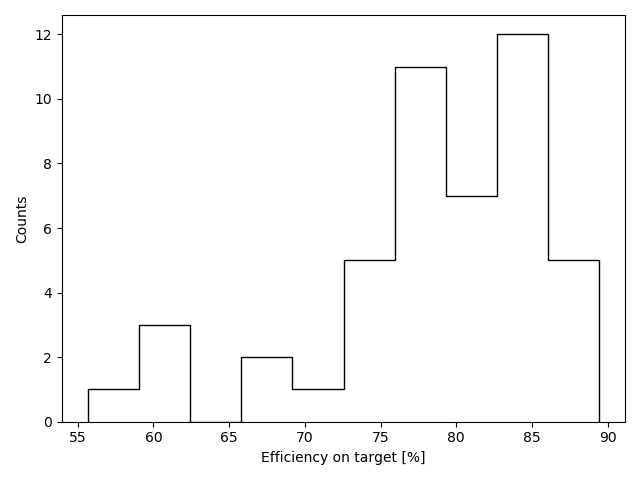}
   \end{tabular}
   \end{center}
   \caption{A quick look at COSMOS field observations. Left: the integrated RMS signal vs integration time for the 637 central KIDs of the HF array; in black are the measurement points, and in red the resulting fit of Eq.~\ref{eq:rms}. Right: the histogram of the mapping efficiency on target as a function of the observing time. The count represents a single consecutive observation session.}
   \label{fig:mapping}
\end{figure}

Figure~\ref{fig:mapping} shows the result of this fit in arbitrary units (a.u.). We found $\beta$=0.51$\pm$0.01 with a Pearson correlation coefficient $\rho=0.998$ and a p-value $\rm{p}<10^{-55}$. The fit with the Eq.~\ref{eq:rms} acts as a continuous quick look at the observations status: the RMS evolves following a $\propto 1/\sqrt{\rm{t}}$ law suggesting that no major issue is highlighted in the observations in terms of overall noise. In addition, Fig.~\ref{fig:mapping} shows the mapping efficiency. We obtain an average efficiency of 79$\pm$8\%, which is calculated as the ratio between the on-source time and the total time dedicated to COSMOS observations. The overheads include scan failures, pointing, telescope slew and focus sessions. This is consistent with the 70\,\% value estimated during the scientific case definition\cite{2020A&A...642A..60C}.

\section{CONCLUSION}
\label{sec:conclusion}

The CONCERTO installation team arrived at APEX on the 6th of April 2021; four days after (the 10th) the first cool-down took place, and the working temperature of 60\,mK was reached just two days after on the 12th of April.
We collected 369\,hours of observation time on the COSMOS project (including overheads) from July 2021 to May 2022, from the ESO, OSO and Chilean allocated time.
The first [CII] LIM survey at high mapping speed with a wide-band low-resolution spectrometer represents a pathfinder experiment and has breakthrough potential in the large-scale evolution of the Universe.

In this paper, we have introduced the CONCERTO instrument and its main characteristics. We have also described and analyzed the main systematic effect that has been found to be related to the vibrations of the beamsplitter membrane. We presented the adopted solution as well as the promising results of its analysis in various configurations.
Finally, we described the mapping strategy as well as the recent result on the observation status of the COSMOS field.

\acknowledgments % equivalent to \section*{ACKNOWLEDGMENTS}       
 
The KID arrays described in this paper have been produced at the PTA Grenoble microfabrication facility. 
CONCERTO project and this work have been supported by the LabEx FOCUS ANR-11-LABX-0013, the European Research Council (ERC) under the European Union's Horizon 2020 research and innovation program (project CONCERTO, grant agreement No 788212), and the Excellence Initiative of Aix-Marseille University-A*Midex, a French ``Investissements d'Avenir'' program.

% References
\bibliography{report} % bibliography data in report.bib

\begin{thebibliography}{10}

\bibitem{choi2020}
{Choi}, S.~K., {Austermann}, J., {Basu}, K., et~al., ``{Sensitivity of the
  Prime-Cam Instrument on the CCAT-Prime Telescope},'' {\em Journal of Low
  Temperature Physics}~{\bf 199},  1089--1097 (Mar. 2020).

\bibitem{crites2014}
{Crites}, A.~T., {Bock}, J.~J., {Bradford}, C.~M., et~al., ``{The TIME-Pilot
  intensity mapping experiment},'' in [{\em
  \procspie}{\nolinebreak\hspace{0.1em}]},  {\em Society of Photo-Optical
  Instrumentation Engineers (SPIE) Conference Series} {\bf 9153},  91531W (Aug.
  2014).

\bibitem{2020A&A...642A..60C}
{Concerto Collaboration}, {Ade}, P., {Aravena}, M., et~al., ``{A wide
  field-of-view low-resolution spectrometer at APEX: Instrument design and
  scientific forecast},'' {\em \aap}~{\bf 642},  A60 (Oct. 2020).

\bibitem{yue2015}
{Yue}, B., {Ferrara}, A., {Pallottini}, A., {Gallerani}, S., and {Vallini}, L.,
  ``{Intensity mapping of [C II] emission from early galaxies},'' {\em
  \mnras}~{\bf 450},  3829--3839 (July 2015).

\bibitem{2013MNRAS.429..849D}
{De Petris}, M., {De Gregori}, S., {Decina}, B., et~al., ``{Atmospheric
  monitoring in the millimetre and submillimetre bands for cosmological
  observations: CASPER2},'' {\em \mnras}~{\bf 429},  849--858 (Feb. 2013).

\bibitem{2018IAUS..333..228L}
{Lagache}, G., ``{Exploring the dusty star-formation in the early Universe
  using intensity mapping},'' in [{\em Peering towards Cosmic
  Dawn}{\nolinebreak\hspace{0.1em}]},  {Jeli{\'c}}, V. and {van der Hulst}, T.,
  eds.,  {\bf 333},  228--233 (May 2018).

\bibitem{2010SPIE.7741E..0MD}
{Doyle}, S., {Mauskopf}, P., {Zhang}, J., et~al., ``{A review of the lumped
  element kinetic inductance detector},'' in [{\em Millimeter, Submillimeter,
  and Far-Infrared Detectors and Instrumentation for Astronomy
  V}{\nolinebreak\hspace{0.1em}]},  {Holland}, W.~S. and {Zmuidzinas}, J.,
  eds., {\em Society of Photo-Optical Instrumentation Engineers (SPIE)
  Conference Series} {\bf 7741},  77410M (July 2010).

\bibitem{fasano-ltd}
{Fasano}, A., {Aguiar}, M., {Benoit}, A., et~al., ``{The KISS Experiment},''
  {\em Journal of Low Temperature Physics}~{\bf 199},  529--536 (Apr. 2020).

\bibitem{fasano_aa}
{Fasano}, A., {Mac{\'\i}as-P{\'e}rez}, J.~F., {Benoit}, A., et~al., ``{Accurate
  sky signal reconstruction for ground-based spectroscopy with kinetic
  inductance detectors},'' {\em \aap}~{\bf 656},  A116 (Dec. 2021).

\bibitem{mpi}
{Martin}, D.~H. and {Puplett}, E., ``{Polarised interferometric spectrometry
  for the millimeter and submillimeter spectrum.},'' {\em Infrared
  Physics}~{\bf 10},  105--109 (1970).

\bibitem{2022JLTP..tmp...51M}
{Monfardini}, A., {Beelen}, A., {Benoit}, A., et~al., ``{CONCERTO at APEX:
  Installation and Technical Commissioning},'' {\em Journal of Low Temperature
  Physics}  (Mar. 2022).

\bibitem{2022EPJWC.25700010C}
{Catalano}, A., {Ade}, P., {Aravena}, M., et~al., ``{CONCERTO at APEX:
  Installation and first phase of on-sky commissioning},'' in [{\em European
  Physical Journal Web of Conferences}{\nolinebreak\hspace{0.1em}]},  {\em
  European Physical Journal Web of Conferences} {\bf 257},  00010 (July 2022).

\bibitem{2017A&A...599A..34R}
{Ritacco}, A., {Ponthieu}, N., {Catalano}, A., et~al., ``{Polarimetry at
  millimeter wavelengths with the NIKA camera: calibration and performance},''
  {\em \aap}~{\bf 599},  A34 (Mar. 2017).

\bibitem{2022ApJS..258...11W}
{Weaver}, J.~R., {Kauffmann}, O.~B., {Ilbert}, O., et~al., ``{COSMOS2020: A
  Panchromatic View of the Universe to z 10 from Two Complementary Catalogs},''
  {\em \apjs}~{\bf 258},  11 (Jan. 2022).

\end{thebibliography}
\bibliographystyle{spiebib} % makes bibtex use spiebib.bst

\end{document}